\documentclass[12pt]{article}
\usepackage[utf8]{inputenc}
\usepackage{comment}
\usepackage{cite}
\usepackage[absolute]{textpos}
\input epsf.sty
\usepackage{setspace}
\usepackage[margin=2cm]{geometry}
\usepackage{verbatim}         
\usepackage{graphicx}						

\usepackage{hyperref}
\usepackage{amssymb}
\usepackage{amsmath}
\usepackage{xcolor}
\hypersetup{
  colorlinks   = true, 
  urlcolor     = blue, 
  linkcolor    = blue, 
  citecolor   = red  
}

\graphicspath{ {./} }

\newcommand{\be}{\begin{eqnarray}\displaystyle}
\newcommand{\ee}{\end{eqnarray}}
\newcommand{\nn}{\nonumber}
\newcommand{\f}{\frac}
\newcommand{\p}{\partial}

\title{Effect of small cosmological constant on\\ electromagnetic memory effect}
\author{}
\date{}

\begin{document}	

\begin{textblock}{5}(6,1)
 \color{red}\Large $||$ Sri Sainath $||$
\end{textblock}

\color{black}

\color{black}
\maketitle

\centerline{\large {  Sayali Atul Bhatkar,  }}

\vspace*{4.0ex}
\centerline{\textit{\large {Tata Institute of Fundamental Research,}}}
\centerline{\small \textit{Homi Bhabha Road, Navy Nagar, Colaba, Mumbai 400005, INDIA.}}
\vspace*{4.0ex}
\centerline{\small E-mail: sayali014@gmail.com.}

\vspace{3cm}
\textbf{Abstract}\\\\
\begin{doublespace}
We consider a generic scattering process that takes place in a region of size $R$ inside the static patch of the de Sitter spacetime such that $R$ is smaller than the curvature length scale of the background. The effect of curvature can thus be studied perturbatively. We obtain the asymptotic electromagnetic field generated by the scattering process including the leading order correction due to the presence of de Sitter background and discuss its universal aspects. We finally caculate the resultant first order corrections to the flat spacetime velocity memory effect.
\end{doublespace}

\newpage

\tableofcontents
\vspace{2cm}

\section{Introduction}

Scattering amplitudes of gauge theories satisfy interesting properties in low energy limit that are encoded in soft theorems \cite{Bloch,soft0,soft1,soft2,soft3,jackiw2,soft4}. In the limit when energy of a scattered photon is taken to 0 (soft), the leading term in scattering amplitudes goes like inverse of the soft energy and the coefficient of this term is a universal soft factor times the lower point amplitude without the soft particle. This soft factor depends only on the electric charge and momenta of the scattering particles and is completely insensitive to the bulk details of the scattering process. This universality has attracted a lot of attention and is known to be a manifestation of an underlying symmetry of the theory \cite{fer1,fer2,qed1,qed2,qed3}. This line of been study has been extended beyond the leading order as well \cite{sub1,sub2}.

Soft theorems have also been studied at the level of classical electromagnetism \cite{G waves, log waves, Sen Laddha}. In the frequency space, the classical field generated by a scattering process exibits universal properties in the low frequency limit. The leading order term in the classical field is universal and goes like inverse of the frequency. The coefficient of this term is equal to the leading order soft factor that appears in quantum soft theorem. This low frequency behaviour controls the late time limit of the classical field. Interestingly, the leading mode in the late time radiative field is observable via the velocity memory effect\cite{mem, mem1,mem2,mem3}. This effect refers to the observable change in the velocity of a free asymptotic charge due to passage of electromagnetic radiation. This change in velocity is controlled by the leading soft factor, it is universal and is insensitive to the bulk details of the scattering process.

Above discussion pertains to electromagnetic radiation emitted in flat spacetime. Since we live in an expanding universe, it is necessary to include the effect of the gravitational background on the above picture. In this paper we aim to discuss the corrections to the velocity memory effect arising due to presence of a small positive cosmological constant ($\Lambda$). Therefore we study the electromagnetic radiation emitted due to scattering of charged particles in de Sitter background. Some features of asymptotically flat spacetimes do not go over to this case. A key feature of gravity with $\Lambda>0$ is that no matter how far one recedes from an isolated body, spacetime curvature does not go to zero. This is different from asymptotically flat spacetimes\cite{Ash}. 
Another peculiar feature of the de Sitter spacetime is that the full spacetime is not observable by any observer. Hence the flat spacetime boundary observables cannot be straightforwardly generalised to this case. In the context of theory of inflation, future boundary correlators in de Sitter spacetime form important observables and have been explored extensively \cite{cor1,cor2}. In \cite{scat}, the authors studied a natural observable associated to a scattering problem (akin to the Minkowski S-matrix) involving evolution of initial data from the past cosmological horizon of the static patch to the future cosmological horizon. We will also consider processes that are confined to a region inside the static patch of the de Sitter spacetime and calculate the radiative field in an appropiately defined late time limit. 

We will treat '$\Lambda$' as a small perturbation avoiding  the technical complexities of full nonlinear gravitational theory. Incorporating the effect of de Sitter background perturbatively, we obtain the first order correction to the late time electromagnetic radiation which in turn is responsible for corrections to the flat spacetime electromagnetic memory effect. Gravitational memory effects have been investigated in de Sitter spacetime\cite{dSmem, dSmem2, dSmem3, dSmem4, dSmem5}. Similar questions have been investigated in the context of Anti de Sitter spacetime in \cite{ads1,ads2} for the special case of classical radiation emitted in probe scatterer approximation. 

We consider a generic scattering process taking place in a region of size $R$ inside the static patch of four dimensional de Sitter spacetime. For validity of the perturbative expansion, we assume that $R$ and all the length scales of the problem like the range of the scattering forces, the distance at which we place the detector, etc are smaller than the curvature length scale of the background $\ell$ ( $\Lambda=\f{3}{\ell^2}$). When any of these length scales become comparable to $\ell$ the non linear effect of background becomes important and the perturbative approximation breaks down. This approximation is applicable to terrestial experiments or also for galactic sources. But due to the extremely small value of $\Lambda$ these effects are inconsequential experimentally. Nonetheless studying these corrections will allow us to explore the underlying symmetries. Therefore this problem has theoretical significance.

The asymptotic radiative field generated by a classical scattering process including the first order correction in $\mathcal{O}(\f{1}{\ell^2})$ is obtained in \eqref{Af}. Then we study the effect of the late time radiation on the motion of an asymptotic test charge which is placed at a large distance $r_0$. Here $r_0$ is a length scale larger than the length scales of the scattering process but smaller than the curvature scale $\ell$. Due to the late time radiation, the velocity of such a charge registers a change of following form 
 \begin{align}
\Big[\Delta W^A\Big]_{\substack{u_0,r_0\rightarrow\infty,\\ |u_0|<r_0<\ell}}\sim\  \frac{1}{ r_0^2} \ \big[\ \mathcal{O}(u_0^0) +  \frac{1}{\ell^2}[r_0 +u_0^2 + u_0 +\mathcal{O}(u^0_0)] \ \big]\  .\nn
\end{align}
Here $u_0$ is a time scale such that $u_0$ is bigger than all the time scales involved in the scattering process but $u_0<r_0$. The $\ell^2$-independent term is controlled by the leading soft factor and gives rise to the flat spacetime velocity memory effect. The rest of the terms are absent in flat space time and arise as a result of the cosmological constant. It turns out that the $u_0^2$ and $u_0$ modes are insensitive to the details of bulk physics and are fixed universally in terms of asymptotic properties of the scattering objects. The coefficients of these modes are given in \eqref{DW}. We have also obtained a part of the $\mathcal{O}(u_0^0)$ mode that depends on the asymptotic trajectories of the scattering particles in \eqref{DW}. We argue that details of the bulk process in principle start affecting at this order but have not calculated such terms explictly.

The outline of the rest of the paper is as follows. In Section 2, we start by rederiving the flat spacetime electromagnetic memory effect and move on to basics of de Sitter spacetime. Revisiting the definition of stereographic co-ordinates, we study the motion of a point particle in these co-ordinates. In Section 3, we derive the Green function for propagation of electromagnetic field on de Sitter background upto $\mathcal{O}(\f{1}{\ell^2})$. In Section 4, we discuss the setup of our scattering problem and use the Green function to derive the asymptotic radiative field generated by the process. Finally in Section 5, we use above results to obtain the corrections to flat spacetime electromagnetic memory effect upto $\mathcal{O}(\f{1}{\ell^2})$. We summarise our results in Section 6.

\section{Preliminaries}
\subsection{Late time radiative field in flat spacetime}

To set up the background for the main calculations, we begin by rederiving the electromagnetic memory term \cite{mem1,mem2,mem3} in the late time radiation emitted in flat spacetime. It is useful to use retarded co-ordinate system to discuss the behaviour of radiation in the far furture. The flat metric takes following form in this co-ordinate system ($u=t-r$) :
\be 
ds^2 = -du^2 - 2dudr + r^2\ 2\gamma_{z\bar{z}}\ dz d\bar{z}; \ \ \gamma_{z\bar{z}} = \frac{2}{(1+z\bar{z})^2}.\nn
\ee
Thus for every value of $(u,r)$ we have a 2-sphere. We will use the unit position 3-vector $\hat{x}$ or $(z,\bar{z})$ interchangeably to describe points on $S^2$. We will often use following parametrisation of a 4 dimensional spacetime point :
\be x^\mu = rq^\mu + u t^\mu,\ \ \  q^\mu=(1,\hat{x}), \ \ \  t^\mu=(1,\vec{0}).\label{q}\ee
$\mu$ takes value from 0 to 3.

In the scattering process, we have some $n'$ number of charged bodies coming in to interact. Let us denote the respective velocities by $V_i^\mu$, charges by $e_i$ and masses by $m_i$ (for $ i=1,\cdots ,n'$). The particles interact for some time $|t|<T$ and $(n-n')$ number of final charged bodies with velocities $V_j^\mu$, charges $e_j$ and masses $m_j$ (for $ j=n'+1,\cdots ,(n-n')$) respectively are produced as a result of the interaction.
Thus the trajectory of an i$^{th}$ incoming particle ($x^\mu_i$) is given by :
$$ x^\mu_i= [V_i^\mu \tau + d_i^\mu]\Theta(-T-\tau).$$
$\tau$ is an affine parameter. We have restricted ourselves to the leading order in coupling $e$ so that we can ignore the effect of long range electromagnetic interactions on the asymptotic trajectories. Similarly, an outgoing particle has the trajectory :
$$ x^\mu_j= [V_j^\mu \tau + d_j]\Theta(\tau-T).$$
The bulk trajectories might have any complicated form depending on the short range forces and will not affect our analysis. The current is given by summing over all particles that participate in the scattering. The asymptotic part of this current can be written down as :
$$j^{\text{asym}}_\sigma(x') = \int d\tau\Big[\sum_{i=n'+1}^n e_i V_{i\sigma}\  \delta^4(x'-x_i)\ \Theta(\tau-T)+\sum^{n'}_{i= 1} e_i V_{i\sigma}\  \delta^4(x'-x_i)\ \Theta(-T-\tau)\Big].$$
Here we have labelled the incoming particles by $i$ running from 1 to $n'$ and outgoing particles by $i$ running from $n'+1$ to $n$. Next we need to find the radiation produced by the above current. In Lorenz gauge, the radiation can be obtained from the equation $\Box A_\mu=-j_\mu$. Using the retarded propagator, we get :
\begin{align}
A^{\text{asym}}_\sigma(x)
&=\frac{1}{2\pi }\int d^4x'\ \delta([x-x']^2)\ j^{\text{asym}}_\sigma(x')\  \Theta(t-t') .\label{A1}
\end{align} 
We have added a superscript to note that we have ignored the bulk sources of radiation. Henceforth, we will drop this superscript but it should be remembered that we are calculating only the asymptotic part of the field. The retarded root of the delta function $\delta([x-x_i(\tau)]^2)$ is given by 
\be \tau_0=-(V_i.x-V_i.d_i)-\sqrt{(V_i.x-V_i.d_i)^2+(x-d_i)^2}.\label{tau0}\ee
Hence, we can write down the total asymptotic field generated by the scattering process. It is given by :
\begin{align}
A_\sigma(x)
&=\sum_{i=n'+1}^n\frac{1}{4\pi }  \frac{e_iV_{i\sigma}\ \Theta(\tau_0-T) }{\sqrt{(V_i.x-V_i.d_i)^2+(x-d_i)^2}}  +\sum_{i=1}^{n'}\frac{1}{4\pi }  \frac{e_iV_{i\sigma}\ \Theta(-T-\tau_0) }{\sqrt{(V_i.x-V_i.d_i)^2+(x-d_i)^2}}\label{A0} \ .
\end{align}
Here $V_i.x$ denotes the Lorenztian dot product between the respective Lorentz vectors.

Next let us take the limit $r\rightarrow\infty$ with $u$ finite in eq.\eqref{A0}. Using $\tau_0= \frac{u}{|q.V_i|}+\mathcal{O}(\frac{1}{r})$ in \eqref{A0} :
\begin{align}
A_\sigma(x)
&=-\frac{1}{4\pi r}\Big[\sum_{i=n'+1}^n  \frac{e_iV_{i\sigma}}{V_i.q} \Theta(u-T) +\sum_{i=1}^{n'}\frac{1}{4\pi }  \frac{e_iV_{i\sigma}}{V_i.q}  \Theta(-T-u) + ...\Big]+\mathcal{O}(\frac{1}{r^2})\ . \label{A00}
\end{align}
The $\frac{1}{r}$-term gives us the radiative field. At large values of $u$, we see that it goes like $u^0$. '$...$' denote $u$-fall offs that are faster than any (negative) power law behaviour. This mode gives rise to a change in velocity of an asymptotic charge. We will calculate the field strength and substitute it in the equation of motion for a point test charge : $m\f{\p V^\mu}{\p\tau}=eF^\mu_\nu V^\mu$, the magnitude of the shift in velocity over a time scale $u_0$ turns out to be
\be m\Delta V^\mu\ \ =\ \ e \int_{-u_0}^{u_0} du\ \p_u A^\mu\ \ =\ \  -\frac{e}{4\pi r}\sum_{i=1}^n \eta_i e_i\frac{ V_i^\mu}{V_i.q} +\mathcal{O}(\f{1}{r^2}). \label{deltaA} \ee
Here $u_0$ is a time scale larger than the time scales of the scattering process ($T$). We have used $\eta_i=1 (-1)$ for outgoing (incoming) particles. Thus the shift is controlled by the leading soft mode as discussed in section III of \cite{mem}. This is called the flat space velocity memory effect \cite{mem1,mem2,mem3}.

To summarise we obtained the asymptotic field generated by a scattering event and discussed its effect on a test charge. As visible from the form of the expression, the amount of kick received by a test charge is insensitive to the details of the scattering. The corrections to \eqref{A00} at higher orders in $e$ have been discussed in \cite{log mem em, 2007} and take following form 
\be A_\mu(x) \sim \ \frac{1}{4\pi r}\Big[ f_{1\mu}(\hat{x}) u^0\ +\ f_{2\mu}(\hat{x}) \f{1}{u}\ +\ f_{3\mu}(\hat{x})\f{\log u}{u^2}\ +\ ...\Big] +\mathcal{O}(\frac{1}{r^2}). \label{falloff}\ee
The $u^0$-mode is uncorrected at  higher orders in $e$ and $f_{1\mu}(\hat{x})$ is given by \eqref{A00}. The corrections are subleading at large $u$. 

\subsection{de Sitter spacetime}
Our goal in this paper is obtain the leading order corrections to \eqref{falloff} arising due to the presence of the cosmological constant and study the resultant corrections to the electromagnetic memory effect. Consider a scattering that takes place in a background with cosmological constant. 
In presence of cosmological constant, the background has to satisfy following equation 
$$R_{\mu\nu}-\f{1}{2}Rg_{\mu\nu}+\Lambda g_{\mu\nu}=0.$$
The maximally symmetric solution to above equation is the well known de Sitter solution. A well known observation made by Schrodinger is that de Sitter spacetime can be embedded in a higher dimensional flat spacetime. Let us start with 5 dimensional flat spacetime. 4 dimensional de Sitter solution can be embedded in this spacetime via the constraint equation \cite{Hawk}
$$ H\ :\  \eta_{AB}X^AX^B=\ell^2.$$
Here $X^A$ are Cartesian co-ordinates in the embedding space, ($A=0,1,...,4$) and $\eta_{AB}$ is the corresponding flat Lorentzian metric. We will use stereographic co-ordinates to descibe de Sitter spacetime. These co-ordinates are defined according to following equations \cite{stereo}
\be x^\mu = \f{2}{1-\f{X^4}{\ell}} X^\mu \ \ ...\ \ (\mu =0,1,2,3).\label{trans}\ee
 The co-ordinate $X^4$ is fixed by the constraint equation and turns out to be\footnote{The other solution to the constraint equation is $X^4=\ell$ where our co-ordinates are ill defined.}
\be X^4=\ell\ \f{{x^2}-4{\ell^2}}{{x^2}+4{\ell^2}}.\label{X4}\ee
In above equation and henceforth $x^2$ is used to denote $\eta_{\mu\nu}x^\mu x^\nu$.  Now we can project the 5 dimensional metric onto the $H$-hyperboloid given by $ \eta_{AB}X^AX^B=\ell^2$, to get the 4 dimensional de Sitter metric. We have $ds^2=\eta_{AB}dX^AdX^B|_H=g_{\mu\nu}dx^\mu dx^\nu$, where
\begin{align}
g_{\mu\nu}&=\Omega^2 \eta_{\mu\nu} ,\ \ \Omega=\f{1}{1+x^2/4\ell^2}.\label{coord}.
\end{align}
Greek indices will be used to denote de Sitter tensors. 

An advantage of using above co-ordinate system is that the metric is conformally flat. We see that $x^2=-4\ell^2$ is a singular surface in this co-ordinate system. But this singularity does not affact our analysis. Our physical set up involves scattering of charged particles in a region of size $R$. This region is such that the points in $R$ have $x^\mu < \ell$ for every component '$\mu$'. Hence $|x^2|<4\ell^2$ for us. Also the transformation in \eqref{trans} is ill defined for $X^4=\ell$. But from \eqref{X4}, we see that our region of interest corresponds to $X^4\sim -\ell$. Hence above co-ordinate system is well defined for the entire region of our interest i.e. region $R$.

Let us discuss where this region of size $R$ sits inside the Penrose diagram of the de Sitter spacetime.
\begin{figure}[h]
\caption{Penrose diagram of the de Sitter spacetime}
\centering
\includegraphics[width=0.62\textwidth]{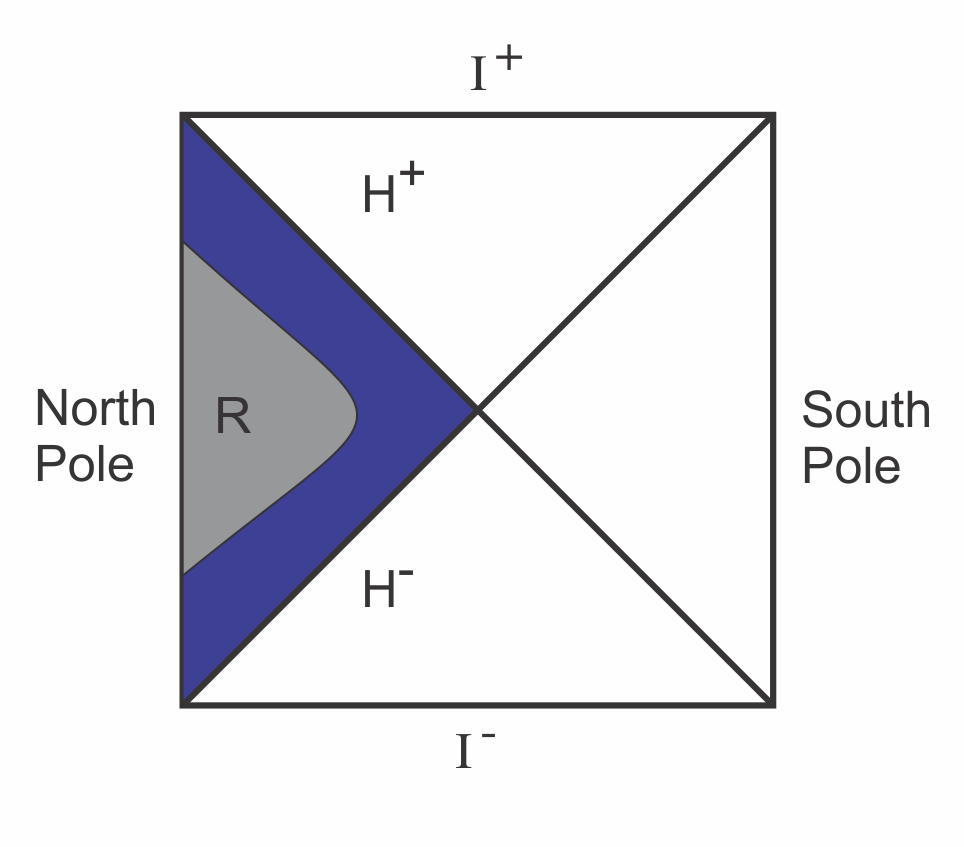}
\end{figure}
The blue coloured triangle in Figure 1 is the static patch of de Sitter spacetime. In terms of the static co-ordinates ($\tilde{t},\tilde{r}, \omega^a$), the stereographic co-ordinates take following form \cite{dSreview}
\be t = -\ell\Omega^{-1} \sqrt{1-\f{\tilde{r}^2}{\ell^2}}\ \text{sinh} \f{\tilde{t}}{\ell}, \ \ x^a = \Omega^{-1}\tilde{r}\omega^a  \  ...\  (a =1,2,3)\ \text{and}\ \ \sum_{a=1}^3(\omega^a)^2=1.\nn\ee
Hence we see that $r< \ell$ and $|t|< \ell$ region is mapped to a region : $\tilde{r}< \ell$ and $|\tilde{t}|< \ell$. Thus the scattering region as well as the detector lie within some region '$R$' in the static patch of the de Sitter spacetime such that $R<\ell$.\\\\
\textbf{Motion of a classical particle in stereographic co-ordinates}\\

Let us study the trajectory of a point particle moving in de Sitter spacetime. We will work in stereographic co-ordinates and expand all the quantities to first order in $\f{1}{\ell^2}$. Hence from \eqref{coord}, we get 
\begin{align}
g_{\mu\nu}&=\Omega^2 \eta_{\mu\nu}=\eta_{\mu\nu}-\frac{x^2}{2\ell^2}\eta_{\mu\nu}+...\ .\label{met}
\end{align}
The inverse metric is given by
\begin{align}
g^{\mu\nu}&=\eta^{\mu\nu}+\frac{x^2}{2\ell^2}\eta^{\mu\nu}+...\ .\nn
\end{align}
The Christoffel's symbols and curvature tensors turn out to be 
\begin{align}
\Gamma^\mu_{\nu\lambda}&=-\frac{1}{2\ell^2}[\delta^\mu_{(\nu} x_{\lambda)}-x^\mu\eta_{\nu\lambda}]+...\ .\nn\\
R_{\mu\nu\lambda\sigma}& =\frac{1}{\ell^2}[\eta_{\mu\lambda}\eta_{\nu\sigma}-\eta_{\mu\sigma}\eta_{\nu\lambda}]+...\ ,\nn\\
R_{\mu\nu}&=\frac{3}{\ell^2}\eta_{\mu\nu}+...\ . \label{christ}
\end{align}

Next we can write down the equation of trajectory for a point particle. Denoting the trajectory of such a particle by $x_i^\mu(\tau)$, the geodesic equation is given by
\begin{align}
\frac{d^2 x^\mu_i}{d \tau^2} + \Gamma^\mu_{\lambda \sigma}\frac{dx^\lambda_i}{d \tau}\frac{dx_i^\sigma}{d \tau}=0.
\end{align}
To the zeroth order in $\frac{1}{\ell^2}$-expansion, the trajectory is simply $$x_i^\mu|_{ \frac{1}{\ell^2}=0}= V_i^\mu \tau + d_i^\mu.$$
The first order correction in $\frac{1}{\ell^2}$ to motion of particle is given by
\begin{align}
\f{d^2x_i^\mu}{d\tau^2}&=-\f{\tau}{2\ell^2}V_i^\mu +\f{V_i^\mu}{\ell^2}V_i.d_i+\f{d_i^\mu}{2\ell^2}.
\end{align}
Above equation is integrated to obtain the velocity. The constant of integration is fixed by demanding $g_{\mu\nu}\f{dx_i^\mu}{d\tau}\f{dx_i^\nu}{d\tau}=-1$. The trajectory turns out to be
\begin{align}
{x_i^\mu}&=V_i^\mu\tau[1+\f{d_i^2}{4\ell^2}-\f{\tau^2}{12\ell^2}+\f{V_i.d_i}{2\ell^2}\tau]+\ d_i^\mu[1+\f{\tau^2}{4\ell^2}].\label{xi}
\end{align}
We consider scattering of freely falling objects. Hence our scattering particles move on trajectories described by above equation 
(both) before and after the scattering event.

\section{The Green's function for Electromagnetic field}
In this section, we solve for the retarded Green's function for minimally coupled U(1) gauge field propagating in de Sitter spacetime upto first order in $\f{1}{\ell^2}$. The equation of motion of the gauge field is given by
$$g^{\mu\nu}\nabla_\mu F_{\nu\sigma} =-j_\sigma.$$ 
Here, $F_{\mu\nu}=\p_{[\mu}A_{\nu]}$, $g^{\mu\nu}$ is the de Sitter metric and $\nabla_\mu$ is the associated covariant derivative. Choosing the generalised Lorenz gauge $\nabla_\mu A^\mu=0$, we get the following equation of motion for the gauge field \cite{spin1}
$$\nabla^2A_\mu - \f{3}{\ell^2} A_\mu = -j_\mu.$$
There is a standard way to calculate the Green's function for the propagation of the U(1) field in arbitrary spacetime described in \cite{EP}. The solution of the gauge field is given by 
\begin{align}
A_\mu(x)  =\f{1}{4\pi}\ g_{\mu\nu}\int d^4x'\ \sqrt{-g'}\ G^{\nu}_{\lambda'}(x,x')\ j^{\lambda'}(x').\label{solu}
\end{align} 
The Green function satisfies the equation 
\begin{align}
\nabla^2 G^{\mu}_{\lambda'}(x,x') -\f{3}{\ell^2}G^{\mu}_{\lambda'}(x,x')\  = -4\pi\delta^4(x,x')\ g^{\mu}_{\lambda'}.\label{G}
\end{align} 
Here $g^\mu_{\ \nu'}$ is the parallel propagator along the geodesic connecting $x$ and $x'$ and $\delta^4(x,x')$ is the covariant Dirac delta function.

The Green function can be obtained using the Hadamard ansatz as discussed in \cite{EP}.
\begin{align}
G^{\mu}_{\lambda'}(x,x') =U^{\mu}_{\lambda'}(x,x')\ \delta_+(\sigma+\epsilon)+V^{\mu}_{\lambda'}(x,x')\ \Theta_+(-\sigma+\epsilon).\label{H}
\end{align}
In above expression $x$ is assumed to lie in the normal convex neighbourhood of $x'$ so that a unique geodesic links these two points. $\sigma$ is the Synge world function which is half the geodesic distance squared between $x$ and $x'$. $\delta_+(\sigma+\epsilon)$ and $ \Theta_+(-\sigma+\epsilon)$ are distributions such that $\delta_+(\sigma+\epsilon)$ has support when $x$ is on the future light cone of $x'$ while $\Theta_+(-\sigma+\epsilon)$ has support when $x$ in the chronological future of $x'$. The small parameter '$\epsilon$' is introduced to make the distributions differentiable at $\sigma=0$ and will be taken to 0 at the end of calculations. The subscript '$+$' refers to the retarded condition i.e. $t>t'$.

Using \eqref{H} and taking $\epsilon\rightarrow 0$, we get
\begin{align}
\nabla^2 G^{\mu}_{\lambda'}(x,x') -\f{3}{\ell^2}G^{\mu}_{\lambda'}(x,x')\ =\ &\ -4\pi\delta^4(x,x')\ U^{\mu}_{\lambda'} \nn\\
 +&\ \   \delta'_+(\sigma)\big[ 2\sigma^\gamma\nabla_\gamma U^{\mu}_{\lambda'}\ + (\sigma^\gamma_\gamma-4)U^{\mu}_{\lambda'}\big]\ + \Theta(-\sigma)\big[\nabla^2 V^{\mu}_{\lambda'}(x,x') -\f{3}{\ell^2}V^{\mu}_{\lambda'}\big]\nn\\
+&\ \   \delta_+(\sigma)\Big[-2\sigma^\gamma\nabla_\gamma V^{\mu}_{\lambda'}\ +\ (2-\sigma^\gamma_\gamma)V^{\mu}_{\lambda'}+\nabla^2 U^{\mu}_{\lambda'}(x,x') -\f{3}{\ell^2}U^{\mu}_{\lambda'}\Big].\label{100}
\end{align} 
Details of derivation of above equation can be found in \cite{EP}. Here we have used $\sigma_\mu$ to denote $\nabla_\mu\sigma$ and $\sigma^\gamma_\gamma$ denotes $\nabla^2\sigma$. Comparing with \eqref{G}, we demand that $U^{\mu}_{\lambda'}\rightarrow g^{\mu}_{\lambda'}$ as $x\rightarrow x'$ and that the last two lines of above equation vanish. This determines $U^{\mu}_{\lambda'}$ and $V^{\mu}_{\lambda'}$ for us. 

Let us  first solve for $U^{\mu}_{\lambda'}$ by setting the coefficient of $ \delta'(\sigma)$ to 0. This has already been done in \cite{EP} and we will quickly review the calculation here. We start with the ansatz $U^{\mu}_{\lambda'}=g^{\mu}_{\lambda'}(x,x') U(x,x')$ where $g^{\mu}_{\lambda'}$ is the parallel propagator and $U(x,x')$ is an arbitrary biscalar. We get following equation for $U(x,x')$
\be 2\sigma^\gamma\nabla_\gamma U\ + (\sigma^\gamma_\gamma-4)U =0. \label{Ueqn}\ee
Next we use a property of the van Vleck determinant. It is defined as follows
\be \Delta=\frac{\text{det}[-\p_\mu\p_\nu'\sigma]}{\sqrt{-g}\sqrt{-g'}}.\label{vD}\ee
The van Vleck determinant satisfies following equation \cite{EP}
$$ (\sigma^\gamma_\gamma-4)=-\sigma^\mu\nabla_\mu\log\Delta=-\f{\p \log\Delta}{\p \tau}.$$
In the last equality we have used $\tau$ to denote the affine parameter on the geodesic joining $x$ and $x'$, so that $\sigma^\gamma\nabla_\gamma =\f{\p }{\p \tau}$. 

Using above expressions in \eqref{Ueqn}, the equation for $U$ takes the form $\f{\p U}{\p \tau} =\f{1}{2}\f{\p \log\Delta}{\p \tau}U$ and we get the solution
\be U^{\mu}_{\lambda'}(x,x')=g^{\mu}_{\lambda'}(x,x')\sqrt{\Delta(x,x')}.\label{U}\ee
It is should be noted that above expression is exact; no approximation has been made so far and in fact it holds for a generic background.

Before proceeding let us briefly discuss the form of field generated by $U^{\mu}_{\lambda'}$. We define
\begin{align}
A^{[I]}_\mu(x)  =\f{1}{4\pi}\ g_{\mu\nu}\int d^4x''\ \sqrt{-g''}\ U^{\nu}_{\lambda'}(x,x'')\ j^{\lambda''}(x'').
\end{align} 
Let us consider a point source with an arbitrary velocity profile $W^\mu(\tau)$ so that the source is given by
\be  j^{\mu}(x)=\frac{e}{\sqrt{-g}}\int d\tau  \ \delta^4(x-x'(\tau))\ W^\mu(\tau). \ee 
$x'(\tau)$ is the trajectory of the source. Integrating over $x''$ we get
\begin{align}
A^{[I]}_\mu(x)  =\f{e}{4\pi}\ g_{\mu\nu}\int d\tau\ g^{\nu}_{\lambda'}(x,x'(\tau))\ \sqrt{\Delta(x,x'(\tau)) }\ \delta_+[\sigma(x,x'(\tau))]\ W^{\lambda'}(\tau).
\end{align} 
Above expression can be written as
\begin{align}
A^{[I]}_\mu(x)  =\f{e}{4\pi}\ \f{\sqrt{\Delta(x,x'(\tau_+)) }}{|\p_\tau \sigma(x,x'(\tau_+)) |}\ g_{\mu\nu'}(x,x'(\tau_+)) \ W^{\nu'}(x'(\tau_+))\ . \label{AI}
\end{align} 
Here $x'(\tau)$ denotes the trajectory of the source and $\tau_+$ is the retarded solution of the equation $\sigma(x,x'(\tau))=0$. Above contribution to the gauge field $A^{[I]}_\mu$ at point $x$ arises due to sources lying on the past light cone of $x$. The gauge field is obtained by parallel transporting velocity $W^{\nu'}$ of the source from the point $x'(\tau_+)$ to the point $x$ along the null geodesic joining the two points and then mutiplying with the factor $\f{\sqrt{\Delta(x,x'(\tau_+)) }}{|\p_\tau \sigma(x,x'(\tau_+))| }$. \\\\\\\\
\textbf{Evaluating \eqref{AI} in de Sitter spacetime in stereographic co-ordinates}.\\\\
Next we will calculate various quantities appearing in \eqref{AI} using the stereographic co-ordinates introduced in \eqref{trans}.
We recall that in \eqref{AI}, $\Delta$ is the van Vleck scalar determinant, $\sigma$ is the Synge function and $g^\mu_{\ \nu'}(x,x')$ is the parallel propgator. We will review the definition of each of these quantities and calculate them explicitly.
The Synge function is given by \cite{EP}
$$\sigma = \frac{1}{2}\int_0^1 d\tau\ g_{\mu\nu} \  \frac{\p \xi^\mu}{\p \tau}\frac{\p \xi^\nu}{\p \tau}.$$
Here $\xi^\mu$ is the geodesic in the de Sitter spacetime such that $\xi^\mu(0)=x'^\mu$ and $\xi^\mu(1)=x^\mu$. We need to calculate $\sigma$ perturbatively to first order in $\frac{1}{\ell^2}$. To the zeroth order we have $\sigma_0=\frac{1}{2}(x-x')^2$. Let $\xi_0^\mu$ denote the geodesic in the flat spacetime hence $\xi_0^\mu = (x^\mu-x'^\mu) \tau + x'^\mu.$ Then the first order correction to the Synge function using \eqref{met} is given by 
\begin{align}
\sigma = \sigma_0\  -\ \eta_{\mu\nu}\ \frac{1}{4\ell^2} \int_0^1 d\tau\  \xi_0^2\ \frac{\p \xi_0^\mu}{\p \tau}\frac{\p \xi_0^\nu}{\p \tau}.\nn
\end{align}
Hence the Synge's function turns out to be
\begin{align}
\sigma(x,x') =& \ \frac{1}{2}(x-x')^2-\frac{1}{4\ell^2} (x-x')^2\big[ x.x'+\f{1}{3} (x-x')^2\big].\label{sigma}
\end{align}
Therefore on a null surface, upto $\mathcal{O}(\frac{1}{\ell^2})$ we get \begin{align}
\p_\tau\sigma&=\ (x'-x).\p_\tau x'[1-\frac{1}{2\ell^2} x.x']\ .\label{sigderi}
\end{align}

We turn to the parallel propagator. Parallel propagator $g^\mu_{\ \nu'}(x,x')$ is defined as a bivector that parallel transports a vector at $x'$ to $x$ along the unique geodesic that links the two points \cite{EP} i.e. 
$$T^\mu(x)=g^\mu_{\ \nu'}(x,x')\ T^{\nu'}(x').$$
The parallel propagator satisfies the equation $\f{\p \xi^\sigma}{\p \tau}\nabla_\sigma g^\mu_{\ \nu'}(\xi,x')=0$, where $\tau$ is an affine parameter on the geodesic $\xi^\sigma(\tau)$ connecting $x$ and $x'$ \cite{CS}. For de Sitter spacetime, upto $\mathcal{O}(\frac{1}{\ell^2})$ the equation of the parallel propagator is as follows
\begin{align}
\frac{\p}{\p\tau}g^\mu_{\ \nu'} &= -\frac{\p \xi_0^\rho}{\p\tau}\ \Gamma_{\nu'\rho}^\mu(\xi_0).\nn
\end{align}
The Christoffel's symbols have been calculated in \eqref{christ}. Since the Christoffel's symbols start at $\mathcal{O}(\f{1}{\ell^2})$ we can use the flat spacetime geodesics $\xi^\mu_0$ in above equation. Above differential equation can be easily solved subject to the boundary condition $g^\mu_{\ \nu'}(x,x)=g^\mu_{\ \nu'}(x',x')=\delta^\mu_{\ \nu'}$. The parallel propagator upto $\mathcal{O}(\frac{1}{\ell^2})$ turns out to be 
\begin{align}
g^\mu_{\ \nu'}(x,x')&= \delta^\mu_{\ \nu'}+\frac{1}{4\ell^2}(x^2-x'^2)+\frac{1}{2\ell^2}(x^\mu\ x'_{\nu'}-x'^\mu\ x_{\nu'}).\label{pp}
\end{align}

Finally we calculate the van Vleck determinant defined in \eqref{vD}. We have calculated the geodesic distance in \eqref{sigma}. Using the same in \eqref{vD}, we get
\begin{align}
\Delta^{1/2}&=1+\frac{(x-x')^2}{4\ell^2}.\label{vV}
\end{align}
Let us recall \eqref{AI}
\begin{align}
A^{[I]}_\mu(x)  =\f{e}{4\pi}\ \f{\sqrt{\Delta(x,x'(\tau_+)) }}{|\p_\tau \sigma(x,x'(\tau_+))| }\ g_{\mu\nu'}(x,x'(\tau_+)) \ W^{\nu'}(x'(\tau_+))\ .\nn
\end{align} 
Using \eqref{sigderi}, \eqref{pp} and \eqref{vV}, we get 
\begin{align}
A^{[I]}_\mu(x)&= \f{e}{4\pi}\ \Big[ \frac{1}{|\p_\tau\sigma|}\eta_{\mu\nu}W^{\nu}(x') [1+\frac{(x-x')^2}{4\ell^2}-\frac{1}{4\ell^2}(x'^2+x^2)\ ]+ \f{1}{2\ell^2}[ x_\mu x'.W - x'_{\mu} x.W]\ \Big]_{\tau=\tau_+}\nn\\
& =\f{e}{4\pi}\ \f{1}{|(x'-x).\p_\tau x' |}\  \Big[\eta_{\mu\nu}W^{\nu}(x')+ \f{1}{2\ell^2}[ x_\mu x'.W - x'_{\mu} x.W]\ \Big]_{\tau=\tau_+}\ .\label{AIf}
\end{align} 
Here $x'(\tau)$ denotes the trajectory of the source and $\tau_+$ is the retarded solution of the equation $\sigma(x,x'(\tau))=0$.
\\\\
\textbf{The tail term in de Sitter spacetime}.\\\\
Next we turn to calculating $V^{\mu}_{\lambda'}(x,x')$. This term leads to propagation of electromagnetic waves inside the null cone. Such terms are called tail terms as they give rise to propagation of signal at speeds less than the vacuum speed of light.

We calculate the tail term perturbatively to first order in $\f{1}{\ell^2}$. Let us revist \eqref{100}. When we demand that the coefficient of $ \delta(\sigma)$ vanish, it fixes $V^{\mu}_{\lambda'}$ in terms of $U^{\mu}_{\lambda'}$ on the null cone via the equation 
$$\text{On }\sigma=0\ \ :\ \ \sigma^\gamma\nabla_\gamma V^{\mu}_{\lambda'}\ +\ \frac{1}{2}(\sigma^\gamma_\gamma-2)V^{\mu}_{\lambda'}=\frac{1}{2}[\nabla^2 U^{\mu}_{\lambda'}(x,x') -\f{3}{\ell^2}U^{\mu}_{\lambda'}].$$
The van Vleck determinant has been calculated in \eqref{vV} to be
$$U^{\mu}_{\lambda'}=g^{\mu}_{\lambda'}(x,x')[1+\f{1}{4\ell^2}(x-x')^2] .$$
Using the expression for parallel propagator in \eqref{pp}, to the first order in $\f{1}{\ell^2}$ we have 
$$\f{1}{2}[\nabla^2 U^{\mu}_{\lambda'}(x,x') -\f{3}{\ell^2}U^{\mu}_{\lambda'}]=-\f{1}{2\ell^2}\delta^{\mu}_{\lambda'} .$$
Thus, we get following equation for $V^{\mu}_{\lambda'}$ 
$$\text{On }\sigma=0\ \ :\ \ \sigma^\gamma\nabla_\gamma V^{\mu}_{\lambda'}\ +\ \frac{1}{2}(\sigma^\gamma_\gamma-2)V^{\mu}_{\lambda'}=-\f{1}{2\ell^2}\delta^{\mu}_{\lambda'}.$$
To leading order, the solution of above equation is given by 
$$V^{\mu}_{\lambda'}=-\f{1}{2\ell^2}\delta^{\mu}_{\lambda'}\ \  \text{on}\ \ \sigma=0.$$
Next we need to calculate $V^{\mu}_{\lambda'}$ inside the null cone. This is done using the third equation obtained from \eqref{100} by setting the coefficient of $\Theta(-\sigma)$ to 0. We have
$$\nabla^2 V^{\mu}_{\lambda'}(x,x') -\f{3}{\ell^2}V^{\mu}_{\lambda'}=0\ \ \text{for}\ \ \sigma<0.$$
Above equation determines $V^{\mu}_{\lambda'}$ inside the light cone. The form of $V^{\mu}_{\lambda'}$ on $\sigma=0$ acts as the boundary data for above equation and allows for a unique solution. 
We can use following ansatz for the bivector $V^{\mu}_{\lambda'}(x,x')$ \cite{vec}
$$V^{\mu}_{\nu'}=-\f{1}{2\ell^2}\delta^{\mu}_{\nu'}[1+f(\sigma)]+\f{1}{\ell^2}g(\sigma)\sigma^\mu\sigma_{\nu'}.$$ 
and we demand that $f,g\rightarrow 0$ as $\sigma\rightarrow 0$.
We need to solve following equation at the leading order : $\Box V^{\mu}_{\lambda'}=0$ where $\Box =-\p_t^2+\vec{\p}^2$ is the flat D Alembertian operator. Using above ansatz for $V^{\mu}_{\nu'}$ we get
$$\sigma^\mu\sigma_{\nu'}\ [g''(\sigma) \sigma^\lambda \sigma_\lambda + g' \sigma^\lambda_\lambda]\ -\ \f{1}{2}\delta^{\mu}_{\nu'}\ [f''(\sigma) \sigma^\lambda \sigma_\lambda + f' \sigma^\lambda_\lambda+4g]\ =\ 0.$$
We can set  the coefficients of the each of these terms to 0. From the first coefficient, we get $g=\f{1}{\sigma}$ but this solution does not have a well defined limit as $\sigma\rightarrow 0$. Hence we are forced to set $g$ to 0. This further leads us to set $f$ to 0 following similar logic.
Thus we get following solution for the tail term
\be V^{\mu}_{\lambda'}=-\f{1}{2\ell^2}\delta^{\mu}_{\lambda'}\ \  \text{for}\ \ \sigma\leq 0. \label{V}\ee
The tail term was missed in \cite{ads1,ads2}.

Let us study the field generated by above part of the Green function. We define
\begin{align}
A^{[II]}_\mu(x)  =\f{1}{4\pi}\ g_{\mu\nu}\int d^4x''\ \sqrt{-g''}\ V^{\nu}_{\lambda'}(x,x'')\ j^{\lambda''}(x'').\nn
\end{align} 
As earlier we consider a point source with an arbitrary velocity profile $W^\mu(\tau)$. It is described by following source 
\be  j^{\mu}(x)=\frac{e}{\sqrt{-g}}\int d\tau  \ \delta^4(x-x'(\tau))\ W^\mu(\tau). \ee 
$x'(\tau)$ is the trajectory of the source. To leading order, we get
\begin{align}
A^{[II]}_\mu(x)  =-\f{e}{8\pi\ell^2}\ g_{\mu\lambda'}\int d\tau\ \Theta_+(-\sigma) \ W^{\lambda'}(\tau).\label{AII}
\end{align} 
Let us compare our results with \cite{Noon}. In \cite{Noon}, the author had argued that in conformally flat spacetimes, the electromagnetic gauge field in general develops a tail term as we see in \eqref{AII}. Also the author had argued that the field strength tensor on the other hand has no tail term in conformally flat spacetimes. Let us calculate the field strength corresponding to \eqref{AII}, we get
\begin{align}
F^{[II]}_{\mu\nu}(x)  =\f{e}{8\pi\ell^2}\ \int d\tau\ \delta_+(\sigma) \ g_{\lambda'[\mu} \p_{\nu]}\sigma\  W^{\lambda'}(\tau).
\end{align} 
Thus consistent with the result of \cite{Noon}, we see that the field strength tensor propagates only on the null cones and contains no tails. It should be noted that \eqref{AII} is not a pure gauge term since it leads to a non-vanishing contribution to the field strength.

\section{Radiative field emitted in a scattering process}
Let us describe our physical setup. The scattering problem involves some $n'$ number of freely falling charged particles coming in. They start around some time $t_0$ and distance $r_0$ such that $r_0, t_0$ are much larger than the length scales involved in the scattering but $r_0, t_0 < \ell$. Let $T$ denote the scale of short range forces that are responsible for scattering. For $t<-T$ the particles continue to fall freely as there are no other forces acting on the particles. The short range forces can be ignored for $|t|>T$. Hence the trajectory of an incoming particle is given by \eqref{xi} :
$$ x^\mu_i= \Big[V_i^\mu \tau[1+\f{d_i^2}{4\ell^2}-\f{\tau^2}{12\ell^2} +\f{\tau}{2\ell^2}V_i.d_i] + d^\mu_i[1+\f{\tau^2}{4\ell^2}]\ \Big]\Theta(-T-\tau).$$
We have denoted the respective velocities by $V_i^\mu$, charges by $e_i$ and masses by $m_i$ (for $ i=1\cdots n'$). $\tau$ is an affine parameter. We have ignored the effect of long range electromagetic interactions on the asymptotic trajectories. The particles come closer and interact in the region $|\tau|<T$. This short range interaction could be of any kind and of any strength.  
$(n-n')$ number of final charged particles with velocities $V_i^\mu$, charges $e_i$ and masses $m_i$ (for $ i=n'+1\cdots (n-n')$) repectively are produced as a result of the interaction. These particles move out. Once they are sufficiently apart ($|\tau|>T$) they continue on geodesics, hence, the outgoing trajectories take the following form
$$ x^\mu_i= \Big[V_i^\mu \tau[1+\f{d_i^2}{4\ell^2}-\f{\tau^2}{12\ell^2}V_i^\mu +\f{\tau}{2\ell^2}V_i.d_i] + d^\mu_i[1+\f{\tau^2}{4\ell^2}]\ \Big]\Theta(\tau-T).$$
It should be noted that this is different from \cite{ads1,ads2} wherein the $\mathcal{O}(\f{1}{\ell^2})$ corrections to trajectories were not considered.

In the asymptotic regime, above process is described by following current
\begin{align}
j^{\mu}(x) =& \sum_{i=n'+1}^n\frac{e_i}{\sqrt{-g}}\int d\tau\  \big[V_i^\mu(1+\f{d_i^2}{4\ell^2})-\f{\tau^2}{4\ell^2}V_i^\mu +[\f{V_i^\mu}{\ell^2}V_i.d_i+\f{d_i^\mu}{2\ell^2}]\tau\big]\  \delta^4(x-x_i)\ \Theta(\tau-T)\nn\\
&+\sum_{i=1}^{n'}\frac{e_i}{\sqrt{-g}}\int d\tau\  \big[V_i^\mu(1+\f{d_i^2}{4\ell^2})-\f{\tau^2}{4\ell^2}V_i^\mu +[\f{V_i^\mu}{\ell^2}V_i.d_i+\f{d_i^\mu}{2\ell^2}]\tau\big]\  \delta^4(x-x_i)\ \Theta(-T-\tau) .\label{j}
\end{align}

Our aim is to obtain the electromagnetic field created by above source including the first order correction due to the cosmological constant. We will substitute above source in Green function solution given in \eqref{solu}. We have seen that the Green function given in \eqref{H} contains two parts : one that is responsible for purely null propagation and a tail part that leads to propagation inside the null cone. The first part has been calculated in \eqref{U} and we also discussed the field generated by this term for a general point source in \eqref{AI}. This expression is evaluated on the retarded root $\tau_+$. The retarded root is obtained by solving the equation $\sigma(x,x_i(\tau))=0$. Since we are working with conformally flat metric given in \eqref{met} null cones continue to be be given by $[x-x_i(\tau)]^2=0$ as seen in \eqref{sigma}. But it should be noted that $x_i(\tau)$ itself contains corrections at $\mathcal{O}(\f{1}{\ell^2})$ since the scattering particles move on geodesics of de Sitter spacetime in the asymptotic region. We can evaluate the root pertubatively. We call the root given in \eqref{tau0} as the zeroth order retarded root. It pertains to the case when scattering particles move on straight line trajectories. So we will evaluate the higher order terms in $[x-x_i(\tau)]^2=0$ on $\tau=\tau_0$. The equation for the corrected root turns out to be 
\begin{align}
&-(x-d_i)^2 +\tau^2+2\tau V_i.(x-d_i)+\frac{1}{\ell^2}\Big[\tau_0[\tau_0+V_i.(x-d_i)]\ [\f{x_{i0}^2}{2}+\f{\tau_0^2}{3}]-\f{1}{2}(x_{i0}-x).d_i\tau_0^2 \Big]=0.\nn
\end{align}
Here we have used $x_{i0}$ to denote the zeroth order trajectory i.e. $x_{i0}^\mu=V_i^\mu\tau_0+d_i^\mu$ to compactify the equation. Above equation is now just a quadratic equation in $\tau$. The retarded root is 
\be \tau_+=-(V_i.x-V_i.d_i)-\Big[X^2+\frac{1}{\ell^2}\big[\ \tau_0X [\f{x_{i0}^2}{2}+\f{\tau_0^2}{3}]+\f{1}{2}(x_{i0}-x).d_i\tau_0^2 \big]\  \Big]^{1/2}.\nn\ee
Here we have used $X$ to denote $X^2=(V_i.x-V_i.d_i)^2+(x-d_i)^2$. We can expand the square root to $\mathcal{O}(\f{1}{\ell^2})$ : 
\be
\tau_+&=&\tau_0-\f{\tau_0}{2\ell^2}[\f{x_{i0}^2}{2}+\f{\tau_0^2}{3}]+\frac{\tau_0^2}{4X\ell^2}(x_{i0}-x).d_i.
\ee
To get a flavour of the correction to the retarded root let us study the asymptotic behaviour of the retarded root. 
\be
\Big[\tau_+\Big]_{\substack{u,r\rightarrow\infty,\\ u<r<\ell}}&=&-\f{u+q.d_i}{q.V_i}-\f{1}{\ell^2}\Big[ \ \f{u^3}{12}\f{1}{(q.V_i)^3} + \mathcal{O}(u^2) \Big]+ \mathcal{O}(\f{1}{r}).\label{taus}
\ee
Since $\f{u^2}{\ell^2}<1$, we see that the leading order behaviour of the retarded root is effectively $\mathcal{O}(u)$. 

Substituting this behaviour in \eqref{AI}, we see that the field at large $r$ and large $|u|$ gets contribution from the sources at large $u$. The bulk source corresponds to region $|\tau|<T$ and contributes to the field at finite values of $u$. Thus to calculate \eqref{AI}, it suffiices to use only the asymptotic trajectories i.e. the source given in \eqref{j}. We will use the velocity profiles given in \eqref{j} and substitute in \eqref{AI} to get the respective contribution to the electromagnetic field. The story is different for the tail term. The field generated due to the tail term was written down in \eqref{AII} and is an integral over the entire region lying inside the past light cone of the field point including the bulk region where scattering takes place. The integral in \eqref{AII} has been done in Appendix \ref{A}. 

Let us collect above terms together. To avoid clutter we first write the field generated by an outgoing particle '$i$'. At the end we will simply sum over all incoming and outgoing particles. Thus using \eqref{j} in \eqref{AI} and \eqref{AIIf}, upto $\mathcal{O}(\f{1}{\ell^2})$ we get
\begin{align}
A^{(i)}_\mu(x)  =\ &\f{e_i}{4\pi}\  \frac{\Theta(\tau-T)}{|(x-x_i).\p_\tau x_i|}  \Big[ V_{i\mu} [1+\f{d_i^2}{4\ell^2}-\frac{\tau^2}{4\ell^2}+\frac{\tau}{\ell^2}V_i.d_i-\frac{\tau}{2\ell^2}x.V_i]+\f{d_{i\mu}}{2\ell^2}(\tau-x.V_i) + \f{1}{2\ell^2}x_\mu x_i.V_i \Big]_{\tau=\tau_+}\ \nn\\
&- \f{e_i}{8\pi\ell^2}\ \Theta(\tau-T)\  [ V_{i\mu}\tau_++d_{i\mu}]\ +\ A_{i\mu}^{\text{bulk}}\ .\label{22}
\end{align} 
Here $V_{i\mu}=\eta_{\mu\nu}V_i^{\nu}$. Similarly all vectors in above expression have been lowered using the flat metric. $A_{i\mu}^{\text{bulk}}$ represents the bulk contribution that cannot be determined without knowing the details of the scattering. In \eqref{AIIf} we have argued that this term starts at $\mathcal{O}(\f{u^0}{r})$. Next we evaluate the denominator of above expression. 
\begin{align}
(x_i-x).\p_\tau x_i
&= -[\tau_++V_i(x-d_i)]\ [1+\f{d_i^2}{4\ell^2}-\f{\tau_0^2}{4\ell^2}+\frac{\tau_0}{\ell^2}V_i.d_i]- x.d_i\f{\tau_0}{2\ell^2}+d_i^2\f{\tau_0}{4\ell^2}+\f{\tau_0^3}{12\ell^2}+\f{V_i.d_i}{4\ell^2}\tau_0^2.\nn
\end{align}
Using above expression we can write down the field generated by by an outgoing particle '$i$' including $\mathcal{O}(\f{1}{\ell^2})$ corrections.
\begin{align}
A^{(i)}_\mu(x)  
=\f{e_i}{4\pi }\ \Theta(\tau_0-T)\ \Bigg[\ & \f{V_{i\mu}}{X} \Big[1-\frac{\tau_0}{2\ell^2}x.V_i+\frac{1}{2X\ell^2}\ \big[x.d_i\tau_0-3\f{\tau_0^2}{2}V_i.d_i-d_i^2{\tau_0}{}-\f{\tau_0^2}{2X}(x_{i0}-x).d_i\ \big]\ \Big]\nn\\
&+\ \frac{1}{2X}\f{d_{i\mu}}{\ell^2}(\tau_0-x.V_i)\ +\ \f{1}{2X\ell^2}x_\mu x_{i0}.V_i\ -\f{1}{2\ell^2}[V_{i\mu}\tau_0+d_{i\mu}]\  \Bigg]+\ A_{i\mu}^{\text{bulk}}.\label{A21}
\end{align} 
Here $X=\sqrt{(V_i.x-V_i.d_i)^2+(x-d_i)^2}$ and $\tau_0$ is the zeroth order retarded root given in \eqref{tau0}. In above expression we will do a gauge transformation to eliminate the term proportional to $x_\mu$.
We will use
$$ \p_\mu f(\tau_0) = -f'(\tau_0)\ \big[\f{1}{X}(x-d_i)_\mu + V_{i\mu}+\f{V_{i\mu}}{X}(V_i.x -V_i.d_i)].$$
For our case $f'(\tau_0)=-x_{i0}.V_i\ \Theta(\tau_0-T)$.
Hence we get 
\begin{align}
A^{(i)}_\mu(x)  
=\f{e_i}{4\pi }\ \Theta(\tau_0-T)\ &\Bigg[\  \f{V_{i\mu}}{X} \Big[1-\frac{V_i.x}{2\ell^2}d_i.V_i+\frac{d_i.V_i}{2\ell^2}x_{i0}.V_i+ \frac{1}{2X\ell^2}\big[x.d_i\tau_0-3\f{\tau_0^2}{2}V_i.d_i-d_i^2{\tau_0}{}\nn\\
&-\f{\tau_0^2}{2X}(x_{i0}-x).d_i \big]\ \Big]+\ \frac{1}{2X}\f{d_{i\mu}}{\ell^2}(V_i.d_i-x.V_i)\  -\f{1}{2\ell^2}V_{i\mu}V_i.d_i-\f{1}{2\ell^2}d_{i\mu} \ \Bigg]\ +\ A_{i\mu}^{\text{bulk}}\ .\label{Ai}
\end{align} 
Similarly we can write down the contribution from incoming particles as well. 

The total field is given by summing over all the particles. 
\begin{align}
A_\mu(x)  
=\f{1}{4\pi }\  \sum_{i=n'+1}^n\Theta(\tau_0-T)\ e_i\ \Bigg[\ & \f{V_{i\mu}}{X} \Big[1-\frac{V_i.x}{2\ell^2}d_i.V_i+\frac{d_i.V_i}{2\ell^2}x_{i0}.V_i+ \frac{1}{2X\ell^2}\big[x.d_i\tau_0-3\f{\tau_0^2}{2}V_i.d_i-d_i^2{\tau_0}{}\nn\\
&-\f{\tau_0^2}{2X}(x_{i0}-x).d_i \big]\ \Big]+\ \frac{1}{2X}\f{d_{i\mu}}{\ell^2}(V_i.d_i-x.V_i)\  -\f{1}{2\ell^2}V_{i\mu}V_i.d_i-\f{1}{2\ell^2}d_{i\mu} \ \Bigg]\nn\\
+\f{1}{4\pi }\  \sum_{i=1}^{n'}\Theta(-\tau_0-T)\ e_i\ \Bigg[\ & \f{V_{i\mu}}{X} \Big[1-\frac{V_i.x}{2\ell^2}d_i.V_i+\frac{d_i.V_i}{2\ell^2}x_{i0}.V_i+ \frac{1}{2X\ell^2}\big[x.d_i\tau_0-3\f{\tau_0^2}{2}V_i.d_i-d_i^2{\tau_0}{}-\nn\\
&-\f{\tau_0^2}{2X}(x_{i0}-x).d_i \big]\ \Big]+\ \frac{1}{2X}\f{d_{i\mu}}{\ell^2}(V_i.d_i-x.V_i)\  -\f{1}{2\ell^2}V_{i\mu}V_i.d_i-\f{1}{2\ell^2}d_{i\mu} \ \Bigg]\nn\\
 +\ & A_{\mu}^{\text{bulk}}\ .\label{Af}
\end{align} 
This is an important result of this paper. It is the expression for the asymptotic field generated by a generic scattering process occuring in region $R$ valid upto $\mathcal{O}(\f{1}{\ell^2})$. Since our expression in \eqref{Af} is covariant, it can be used to obtain the field in any co-ordinate system. Next we will use above expression to derive the corrections to the memory effect.

\section{Corrections to the velocity memory effect}
Let us consider a test charge ($e$). At early times there is no electromagnetic field : $F_{\mu\nu}=0$. So the charge moves on a geodesic and satisfies following equation of motion $m\frac{\p W^\mu}{\p \tau}+\Gamma^\mu_{\nu\lambda}W^\nu W^\lambda=0$. $W^\mu = \frac{\p x^\mu}{\p \tau}$ is the covariant velocity of the test charge. The charge is placed at a very large distance $r=r_0$  away from the scattering event but $r_0<\ell$. The velocity can be written as $W^\mu = W^0(1,\vec{w})$. Since the particle is massive we have $|\vec{w}|<1$ where we have set the vacuum speed of light to be 1. The spatial acceleration given by $\Gamma^a_{\nu\lambda}W^\nu W^\lambda$ for $a=1,2,3$ is of $\mathcal{O}(|\vec{w}|)$. If we consider the case $|\vec{w}|<<1$, we can ignore the gravitational acceleration experienced by the test charge.

After the scattering process takes place the emitted radiation eventually reaches the test charge  and it will experience electromagnetic force. The equation of motion is given by
\begin{align}
m\frac{\p W^\mu}{\p \tau}=eF^{\mu}_{\ \nu}W^\nu.
\end{align}
The electromagnetic interection has a component which is $\mathcal{O}(|\vec{w}|^0)$. Due to our approximation of $|\vec{w}|<<1$, we can ignore the higher order terms. Since we assume the magnitude of the 3-velocity is small we effectively ignore the magnetic part of the interaction. The position of the test charge takes following form
$$ r = r_0 + |\vec{w}| t + \mathcal{O}(e)+ \mathcal{O}(\f{1}{\ell^2}).$$
As discussed earlier $r_0$ is larger than all length scales of the problem except the curvature length. Hence it suffices to approximate $r=r_0$ in all our expressions. Next we turn to the components of the velocity that are transverse to the radial direction. Working in the $\lbrace r,u,z^A\rbrace$ co-ordinates defined in \eqref{q}, we have $W^A=\f{1}{r}\p^Aq_\mu W^\mu$.
\begin{align}
m\frac{\p W^A}{\p \tau}=eF^{A}_{\ u}W^u\ \Rightarrow\ m\frac{\p W^A}{\p u}=eF^{A}_{\ u}.\label{test}
\end{align} 
The field strength tensor component is as follows : 
\begin{align}
F^{A}_{\ u}  &=\ \f{1}{r^2}\ [1-\f{ur}{\ell^2}-\f{u^2}{2\ell^2}]\ \gamma^{AB}\ F_{Bu}.\ 
\end{align} 
We calculate the field strength tensor in \eqref{Fmunu}. Also we recall that $F_{Bu}=  r (\p_Bq^\mu)F_{\mu t}$ via co-ordinate transformation given in \eqref{q}. Let us consider the following terms that arise from '$ [\f{ur}{\ell^2}+\f{u^2}{2\ell^2}]F_{Bu}$' :
\begin{align}
-[\f{ur}{\ell^2}+\f{u^2}{2\ell^2}]\ \Big[ \sum_{i=n'+1}^ne_i\f{\p_B q.V_i}{q.V_i}\ \p_u\Theta(u-T) + \sum_{i=1}^{n'}e_i\f{\p_B q.V_i}{q.V_i}\ \p_u\Theta(-u-T)\ \Big].\nn
\end{align}
These terms are of the form $u\delta(u-T)$ or $u^2\delta(u-T)$. Thus these terms have support in $u\sim T$ region and are sensitive to bulk details. Such terms are similar to the bulk terms we had in \eqref{Af}. It is strange that the $\mathcal{O}(\f{r}{\ell^2})$ term also depends on the bulk. Using \eqref{Fmunu}, the field strength can be written as 
\begin{align}
F^{A}_{\ u}=\f{\gamma^{AB}}{r^2}\ [1-\f{ur}{\ell^2}]&\ \p_u\Bigg[ \f{1}{4\pi }\Big[\sum_{i=n'+1}^ne_i\f{\p_B q.V_i}{q.V_i}\ \Theta(u-T) + \sum_{i=1}^{n'}e_i\f{\p_B q.V_i}{q.V_i}\ \Theta(-u-T)\Big]\ \nn\\
&-\f{1}{8\pi }\f{u^2}{\ell^2}\ \Big[ \sum_{i=n'+1}^ne_i\f{\p_B q.V_i}{(q.V_i)^3}\ \Theta(u-T) + \sum_{i=1}^{n'}e_i\f{\p_B q.V_i}{(q.V_i)^3}\ \Theta(-u-T)\ \Big]\nn\\
&-\f{1}{8\pi }\ \f{u}{\ell^2}\ (\p_B q^\mu)q^\sigma\ \Big[  \sum_{i=n'+1}^n\frac{e_i \ J_{i\mu\sigma} }{(q.V_i)^3} \     \Theta(u-T) + \sum_{i=1}^{n'}\frac{e_i \ J_{i\mu\sigma} }{(q.V_i)^3}\   \Theta(-u-T) \Big]\nn\\
&-\f{1}{8\pi }\ \f{1}{\ell^2}\ (\p_B q^\mu) q^\sigma\ \Big[  \sum_{i=n'+1}^n\frac{e_i \ J_{i\mu\sigma} }{(q.V_i)^3}\ q.d_i\  \Theta(u-T) +  \sum_{i=1}^{n'}\frac{e_i  \ J_{i\mu\sigma}}{(q.V_i)^3} \ q.d_i\ \Theta(-u-T) \Big]\ \Bigg] \nn\\
&+\text{bulk terms}+\ \ \mathcal{O}(\f{1}{r^3})\ .
\end{align} 
We reiterate that 'bulk terms' in above expression refer to $\ell$ dependent corrections to $u^0r^0$ mode. We use above expression in \eqref{test} and integrate the equation over a time scale $u_0$ which is greater than the time scale of the scattering process ($T$) but smaller than $r_0$. We get for the tranvserse components of the velocity 
\begin{align}
&\ \ \ \gamma_{AB}r_0^2 \Big[m \Delta W^A\Big]_{\substack{|u_0|,r_0\rightarrow\infty,\\ |u_0|<r_0<\ell}}\nn\\
&=\f{e}{4\pi }\ [1-\f{r_0T}{\ell^2}] \sum_{i=1}^n\eta_ie_i\f{\p_Bq.V_i}{q.V_i}
-\f{e}{8\pi }\ \f{u_0^2}{\ell^2}\ \sum_{i=1}^n\eta_ie_i \frac{ \p_Bq.V_i}{(q.V_i)^3}\nn\\ 
&-\f{e}{8\pi }\ \f{u_0}{\ell^2}\ \sum_{i=1}^ne_i \frac{ (\p_B q^\mu)q^\sigma }{(q.V_i)^3}\ J_{i\mu\sigma} \  -\f{e}{8\pi }\ \f{1}{\ell^2}\ \sum_{i=1}^n\eta_ie_i \ q.d_i\ \frac{(\p_B q^\mu)q^\sigma }{(q.V_i)^3} \ J_{i\mu\sigma} \ +b_B\ .\label{DW}
\end{align} 
We recall that $e_i, V_i^\mu$ are respectively the charges and asymptotic velocities of the scattered particles. $ J_{i\mu\nu} =  d_{i\mu}V_{i\nu}-  V_{i\mu}d_{i\nu}$ is the orbital angular momentum of the $i^{th}$ particle. $\eta_i=1 (-1)$ for outgoing (incoming) particles. $q^\mu$ defined in \eqref{q} captures the angular position of the field point.

The first ($\ell^2$-independent) term in above expression is the so called electromagnetic velocity memory effect. We obtain the corrections to this effect as result of the cosmological constant. The corrections are of $\mathcal{O}(\f{r_0}{\ell^2}), \mathcal{O}(\f{u^2_0}{\ell^2}), \mathcal{O}(\f{u_0}{\ell^2})$ and $\mathcal{O}(\f{1}{\ell^2})$. The $\mathcal{O}(\f{u^2_0}{\ell^2})$ mode is universally fixed in terms of the asymptotic momenta and charges of the scattering particles. We expect that this mode is insensitive to other attributes of the scattering objects. Hence even for the case of scattering of bodies of a finite size (say with some non uniform charge distribution) the coefficient of this mode is expected to remain unchanged. The $\mathcal{O}(\f{u_0}{\ell^2})$-mode depends on the orbital angular momentum of the particles. So we expect that this mode would get modified by the intrinsic spin of the scattering objects and also in presence of non-minimal electromagnetic coupling. Nonetheless these modifications are expected to be universal and insensitive to the details of the scattering. On the other hand the $\mathcal{O}(\f{1}{\ell^2})$-mode is expected to be have non universal terms (denoted by $b_B$) that depend on such details. We have extracted out the 'asymptotic' part of this mode in \eqref{DW}. The $\mathcal{O}(\f{r_0}{\ell^2})$ term also depends on the details of the bulk and has not been calculated explicitly.

The structure of the universal coefficients is reminiscent of the soft factors present in flat spacetime. This calls for a thorough examination of existence of symmetry underlying the new modes.

\section{Summary}
Recent investigations have shed light on rich structure of IR physics of gauge
theories and gravity around flat spacetime. One of the significant outcome of this analysis is the existence of classically observable velocity memory effect due to passage of electromagnetic radiation.

In this paper we have obtained the first order corrections to the flat spacetime electromagnetic memory effect arising due to presence of a de Sitter background. We have considered a generic scattering process taking place in a region of size $R$ inside the static patch
of the de Sitter spacetime such that $R<\ell$. This allows us to study the emitted radiation perturbatively in $\f{1}{\ell^2}$. Our process involves scattering of freely falling particles and is described by the current given in \eqref{j}. As seen in this expression we do not assume anything about the bulk details of the scattering process. We calculate the asymptotic radiative field generated by such a classical scattering process i.e. the field at distance $r_0$ much larger than the scattering length scale but smaller than the curvature length scale $\ell$. This expression has been obtained in \eqref{Af} and includes the first order correction in $\mathcal{O}(\f{1}{\ell^2})$. 

We analysed the effect of the late time radiation on the motion of an asymptotic test charge. Due to the late time radiation, the velocity of such a charge registers a shift that has been calculated in \eqref{DW}. We have already discussed the universal aspects of this expression and will not repeat it here. The most interesting aspect of this result is the close resemblance of the universal coefficients to the flat spacetime soft factors. This hints that the new modes arising in presence of a small cosmological constant could be controlled by underlying symmetries. We leave the investigation of these (perturbative) symmetries to the future. 

It should be emphasised that our setup is different from \cite{dSmem,dSmem2,dSmem3} that study gravitational memory in dS spacetime. \cite{dSmem} studied souces of the special form as given in eqn (31) of that paper. The final result of \cite{dSmem} given in eqn (69) of the paper involves $F\times (1+rH)$ where $(1+rH)$ is the redshift factor and $F$ is the flat space memory term and is given as integral of the flat space source ($\int du\ L$ in the notation of \cite{dSmem}). In this paper our physical event is described by the source given in \eqref{j} such that the scattered particles move on geodesics of de Sitter spacetime which is different from flat spacetime sources. Similar difference is also seen in the approach of \cite{dSmem3}. Quoting the results of \cite{dSmem3} : "we have shown that if we identify the FLRW spacetime with Minkowski spacetime via the coordinates (7) in such a way that $a(\eta_s) = 1$, and we place the same physical source at q and the same physical detector at p in both spacetimes, then the memory effect in the FLRW spacetime will be a factor of $\f{1}{a(\eta_s)} = \f{1}{(1 + z)}$ smaller than the corresponding
memory effect in Minkowski spacetime." This is different from the class of scattering events that we consider.

Let us conclude with some futher questions that can probed in this context. It would be interesting to include the effect of long range electromagnetic interactions between the scattered charges on the late time radiation. We believe that the $u_0^2$ mode in \eqref{DW} should be uncorrected by long range forces. It can be argued that the long range interactions would give rise to a universal $\mathcal{O}(u_0\log u_0)$ mode in \eqref{DW}. Another question to probe is : how do the gravitational interactions between the scattering particles affect the velocity shift given in \eqref{DW}? 

An intiguing question that needs to be understood is the nature of bulk corrections to \eqref{DW}. It is likely that the bulk correction to the $\mathcal{O}(\f{1}{\ell^2})$-mode is the analogue of the remainder terms present in flat spacetime case \cite{inf1,inf}. In this paper we have considered a generic scattering process. Instead one can consider a particular scattering process such that the bulk trajectories are also completely fixed (and known) and then determine the bulk corrections to \eqref{DW}.  It should be worth exploring the structure of these bulk terms to compare it with the structure of the flat spacetime remainder terms.

\vspace{1cm}

\section{Acknowledgements}
I am extremely thankful to Nabamita Banerjee, Arindam Bhattacharjee, Arpita Mitra and Amitabh Virmani for the discussions. I am deeply grateful to my family for their constant support without which this work would not have been possible.

\appendix

\section{Contribution of the tail term}\label{A}
Substituting the source \eqref{j} in the tail term \eqref{AII}, we get the contribution of the tail term to be 
\begin{align}
A^{[II]}_\mu(x)  =-\sum_{i=n'+1}^n\f{e_iV_{i\mu}}{8\pi\ell^2}\ \int_T^\infty d\tau\ \Theta_+(-\sigma) -\sum_{i=1}^{n'}\f{e_iV_{i\mu}}{8\pi\ell^2}\ \int_{-\infty}^{-T} d\tau\ \Theta_+(-\sigma) \ -\f{1}{8\pi\ell^2} \int_{|t'|<T} d^4x'\ \Theta_+(-\sigma)\ j^{\text{bulk}}_\mu(x').
\end{align} 
Here $V_{i\mu}=\eta_{\mu\lambda}V_i^\lambda$. We recall that $\sigma$ is half the geodesic distance i.e. $\sigma=\f{1}{2}[x-x_i(\tau)]^2$. The superscript '$+$' refers to the condition that $t>t'$ i.e. $t$ has to lie in future of the source. The last term represents the contribution of the bulk source. The explicit form of the bulk current depends on the details of the scattering process. We will not assume anything about the bulk details. It should be noted that the second integral diverges at lower limit. We will see that this divergence does not affect any physical quantities. Let us regulate this integral with a lower limit '$L$'.\\

We are interested in the asymptotic field i.e. at large $r$ and large $|u|$. 
Now $\Theta_+(-\sigma)\ \Rightarrow \ \tau_0 -\tau >0$. Using $\tau_0 \sim \f{u}{|q.V_i|} +...$, we get 
\be
\Big[A^{[II]}_\mu\Big]_{\substack{u>>T}}&=&-\sum_{i=n'+1}^n\f{e_iV_{i\mu}}{8\pi\ell^2}\ [\tau_0-T]\  -\sum_{i=1}^{n'}\f{e_iV_{i\mu}}{8\pi\ell^2}\ [-T-L]  \ -\f{1}{8\pi\ell^2} \int_{|t'|<T} d^4x'\  j^{\text{bulk}}_\mu(x')\nn\\
\Big[A^{[II]}_\mu\Big]_{\substack{u<<-T}}&=&  -\sum_{i=1}^{n'}\f{e_iV_{i\mu}}{8\pi\ell^2}\ [\tau_0-L]  \ .
\ee
Above expression can be rewritten as 
\be
\Big[A^{[II]}_\mu\Big]_{\substack{|u|,r\rightarrow\infty,\\ |u|<r<\ell}}&=&-\sum_{i=n'+1}^n\f{e_iV_{i\mu}}{8\pi\ell^2}\ \tau_0\ \Theta(\tau_0-T)  -\sum_{i=1}^{n'}\f{e_iV_{i\mu}}{8\pi\ell^2}\ \tau_0\  \Theta(-T-\tau_0)+\sum_{i=1}^n\f{e_iV_{i\mu}}{8\pi\ell^2}\  T\ \Theta(\tau_0-T) \nn \\ 
&-& \ \f{1}{8\pi\ell^2} \int_{|t'|<T} d^4x'\  j^{\text{bulk}}_\mu(x')\ \Theta(\tau_0-T)\ + \sum_{i=1}^{n'}\f{e_iV_{i\mu}}{8\pi\ell^2}\ L \ . \label{A11}
\ee
Thus, we see that the last term which represents the diverging piece is a constant and will drop out of the field strength tensor and other physical quantities. 

Next let us try to evaluate relevant part of the bulk piece. We will use a trick \cite{Sen Laddha, 2101} that exploits the conservation law of the U(1) current. Using $\p_\mu \sigma = x_\mu -x'_\mu$, we get 
$$x^\mu A^{\text{bulk}}_\mu =  -\f{1}{8\pi\ell^2}\int_{|t'|<T} d^4x'\  j^{\text{bulk}}_\mu(x')\ \p^\mu \sigma+\mathcal{O}(r^0).$$
Now we know that $ \p^\mu \sigma=- \p'^\mu \sigma.$ Further using conservation of current, we are left with boundary pieces
$$x^\mu A^{\text{bulk}}_\mu = -\f{1}{8\pi\ell^2}\int_{|t'|=T} d^4x'\ t^\mu j^{\text{bulk}}_\mu(x')\ \sigma(x,x')\ sign(t')\ +\mathcal{O}(r^0).$$
$t^\mu=\lbrace 1,\vec{0}\rbrace$. At $|t'|=T$ we can use the form of current given in \eqref{j} which is valid for $|t'|>T$. Then it is simple to do the $x'$ integral and we get 
$$x^\mu A^{\text{bulk}}_\mu =  \f{1}{16\pi\ell^2}\sum_{i=1}^n \eta_ie_i\ x^\mu[x_\mu-2\eta_iV_{i\mu}T-2d_{i\mu}]+\mathcal{O}(r^0).$$
$\eta_i =1$ for outgoing particles and $\eta_i=-1$ for incoming particles. The first term in the square bracket vanishes due to conservation of charge and we get 
$$A^{\text{bulk}}_\mu =-  \f{1}{8\pi\ell^2}\Big[\sum_{i=1}^n \eta_ie_i [\eta_iV_{i\mu}T+d_i]+\mathcal{O}(\f{u^0}{r})\Big].$$
Using above expression in \eqref{A11}, we get after dropping the unphysical constant mode
\be
\Big[A^{[II]}_\mu\Big]_{\substack{|u|,r\rightarrow\infty,\\ |u|<r<\ell}}&=&-\sum_{i=n'+1}^n\f{e_iV_{i\mu}}{8\pi\ell^2}\ \tau_0\ \Theta(\tau_0-T)  -\sum_{i=1}^{n'}\f{e_iV_{i\mu}}{8\pi\ell^2}\ \tau_0\  \Theta(-T-\tau_0)\nn \\ 
&-&  \f{1}{8\pi\ell^2}\sum_{i=1}^n \eta_ie_i \ d_{i\mu}\ \Theta(\tau_0-T)+\mathcal{O}(\f{u^0}{r}). \label{AIIf}
\ee

\section{The field strength}
Let us calculate the expression of the field strength. From \eqref{Af}, we have 
\begin{align}
A_\mu(x)  
=\f{1}{4\pi }\  \sum_{i=n'+1}^n\Theta(\tau_0-T)\ e_i\ \Bigg[\ & \f{V_{i\mu}}{X} \Big[1+ \frac{1}{2X\ell^2}\big[x.d_i\tau_0-3\f{\tau_0^2}{2}V_i.d_i-d_i^2{\tau_0}{}-\f{\tau_0^2}{2X}(x_{i0}-x).d_i \big]\ \Big]+\ \frac{1}{2X}\f{d_{i\mu}}{\ell^2}\tau_0\   \Bigg]\nn\\
+\f{1}{4\pi }\  \sum_{i=1}^{n'}\Theta(-\tau_0-T)\ e_i\ \Bigg[\ & \f{V_{i\mu}}{X} \Big[1+ \frac{1}{2X\ell^2}\big[x.d_i\tau_0-3\f{\tau_0^2}{2}V_i.d_i-d_i^2{\tau_0}{}-\f{\tau_0^2}{2X}(x_{i0}-x).d_i \big]\ \Big]+\ \frac{1}{2X}\f{d_{i\mu}}{\ell^2}\tau_0\   \Bigg]\nn\\
 +\ & A_{\mu}^{\text{bulk}}\ .
\end{align} 

\begin{align}
F_{\mu\nu}(x)  
&=\f{1}{4\pi r}\  \sum_{i=n'+1}^n e_i\ \p_u\Big[\ \Theta(\tau_0-T)\   \f{q_{[\mu}V_{i\nu]}}{q.V_i} \Big[1- \frac{\tau_0\ q.d_i}{2\ell^2q.V_i}\ \Big]+\Theta(\tau_0-T)\ \frac{1}{2}\f{q_{[\mu}d_{i\nu]}}{\ell^2q.V_i}\tau_0\   \Big]\nn\\
&+\f{1}{4\pi r }\  \sum_{i=1}^{n'}e_i\ \p_u\Big[\ \Theta(-T-\tau_0)\   \f{q_{[\mu}V_{i\nu]}}{q.V_i} \Big[1- \frac{\tau_0\ q.d_i}{2\ell^2q.V_i} \Big]+\Theta(-T-\tau_0)\ \frac{1}{2}\f{q_{[\mu}d_{i\nu]}}{\ell^2q.V_i}\tau_0\   \Big]\nn\\
 &+\ \ \ A_{\mu}^{\text{bulk}}\ \ \ +\ \ \mathcal{O}(\f{1}{r^2})\ \ .\label{Fmunu}
\end{align}

\end{document}